\documentclass{emulateapj}
\usepackage{amssymb}
\usepackage{amsmath}

\newif\ifAMStwofonts
\AMStwofontstrue

\def\ga{\mathrel{\hbox{\rlap{\hbox{\lower4pt\hbox{$\sim$}}}\hbox{$>$}}}}
\def\la{\mathrel{\hbox{\rlap{\hbox{\lower4pt\hbox{$\sim$}}}\hbox{$<$}}}}

\newcommand{\hcop}{HCO$^+$}
\newcommand\pcc{\mbox{${\rm cm^{-2}}$}}

\newcommand{\hcopwidth}{$\Delta v=4.5\pm 1.0\rm\,km\,s^{-1}$}
\newcommand{\hcopvel}{$v=214.0 \pm 0.4\rm\,km\,s^{-1}$}
\newcommand{\hcopdepth}{$\tau(\rm HCO^+)=0.10\pm 0.02$}
\begin{document}

\title{First Detection of \hcop Absorption in the Magellanic System}

\author{Claire E. Murray$^{\dagger}$\altaffilmark{1},
        Sne\v{z}ana Stanimirovi\'{c}\altaffilmark{1}, 
        N.~M. McClure-Griffiths\altaffilmark{2}, 
        M.~E. Putman\altaffilmark{3},
        H.~S. Liszt\altaffilmark{4},
        Tony Wong\altaffilmark{5},
        P. Richter\altaffilmark{6,7},
        J.~R. Dawson\altaffilmark{8,9},
        John M. Dickey\altaffilmark{10}, 
        Robert R. Lindner\altaffilmark{1}, 
        Brian L. Babler\altaffilmark{1}, 
        J.~R. Allison\altaffilmark{9}
        }

\altaffiltext{$\dagger$}{cmurray@astro.wisc.edu}
\altaffiltext{1}{Department of Astronomy, University of 
                 Wisconsin - Madison, WI 53706, USA}
\altaffiltext{2}{Research School for Astronomy \& Astrophysics,
		Mount Stromlo Observatory,
		Cotter Road, Weston Creek, ACT 2611, Australia}
\altaffiltext{3}{Department of Astronomy, 
		Columbia University, New York,
		NY 10027, USA}
\altaffiltext{4}{National Radio Astronomy Observatory, 
		520 Edgemont Road, Charlottesville,
		VA 22903-2475, USA}
\altaffiltext{5}{Department of Astronomy, University of Illinois, 
		1002 West Green Street, Urbana, IL 61801, USA}
\altaffiltext{6}{Institut f\"ur Physik und Astronomie, 
		Universit\"at Potsdam, Karl-Liebknecht-Strasse 24/25, 
		14476 Potsdam-Golm, Germany}
\altaffiltext{7}{Leibniz-Institut f\"ur Astrophysik Potsdam (AIP), 
		An der Sternwarte 16, 14482 Potsdam, Germany}
\altaffiltext{8}{Department of Physics and Astronomy
 		and MQ Research Centre in Astronomy, 
		Astrophysics and Astrophotonics, 
		Macquarie University, NSW 2109, Australia}
\altaffiltext{9}{CSIRO Astronomy and Space Science, 
		P.O. Box 76, Epping, NSW 1710, Australia} 
\altaffiltext{10}{University of Tasmania, School of Maths 
                 and Physics, Private Bag 37, Hobart, 
                 TAS 7001, Australia}            

\begin{abstract}

We present the first detection of \hcop\ absorption 
in the Magellanic System. Using the Australia Telescope Compact Array (ATCA), 
we observed 9 extragalactic radio continuum sources behind the Magellanic System
and detected \hcop\ 
absorption towards one source located behind the 
leading edge of the Magellanic Bridge. The detection is 
located at LSR velocity of \hcopvel, with a full width at half 
maximum of \hcopwidth\ and optical depth of \hcopdepth. 
Although there is abundant neutral hydrogen (H\,{\sc i}) surrounding 
the sightline in position-velocity space, at the exact location of 
the absorber the H\,{\sc i} column density is low, $<10^{20}\rm\,cm^{-2}$, 
and there is little evidence for dust or CO emission from \emph{Planck} observations. 
While the origin and survival of molecules in such a diffuse environment remains unclear,
dynamical events such as H\,{\sc i} flows and cloud collisions in this interacting system 
likely play an important role.

\end{abstract}


\section{Introduction}

\begin{table*}
\caption{Source Information}
\centering
\begin{tabular}{lcccccccc}
\hline
\hline
Source     & RA (J2000) & Dec (J2000) & $l$         & $b$         & $S_{3\rm\,mm}$  & Time  & $\sigma_{\tau(\rm HCO^+)}$  & Type$^b$ \\
                  & (hh:mm:ss)  & (dd:mm:ss)    & ($^{\circ}$) & ($^{\circ}$) & (Jy)                         & (hr)  	         &  (per $0.8\rm\,km\,s^{-1}$)$^a$    &     \\
\hline
J0056-572    & 00:58:46.6     & -56:59:11.5 &       300.926 &      -60.113  & 0.9 & 2.7   & 0.059    &     \\ %ATCA
J0102-7546 & 01:02:18.8     & -75:46:53.0 &       302.043 &      -41.328   &  0.6 & 1.7  &  0.090  &   AT20G \\ %AT20G
J0208-512    & 02:10:46.2     & -51:01:01.9 &       276.102 &      -61.778   &  2.1 & 2.3  & 0.051    &     \\ %ATCA
J0311-7651 & 03:11:55.3     & -76:51:51.0 &       293.440 &      -37.553    &  0.8 & 2.3  & 0.054     &   AT20G    \\ %AT20G
J0440-6952 & 04:40:47.8     & -69:52:18.1 &       281.836 &      -36.399   &  0.6 & 4.0  &  0.046   &     \\ %ATCA
J0454-810    & 04:50:05.4     & -81:01:02.2 &       293.851 &      -31.371   &  1.5 & 3.3  &  0.026   &      \\ %ATCA
J0506-6109 & 05:06:44.0     & -61:09:41.0 &       270.550 &      -36.072    & 0.4 & 1.7  &   0.090    &   AT20G   \\ %AT20G
J0530-727    & 05:29:30.0     & -72:45:28.5 &       283.850 &      -31.857  & 0.8 & 3.3   &  0.019    &    \\ %ATCA
J0637-752    & 06:35:46.5     & -75:16:16.8 &       286.368 &      -27.158   & 1.8 & 2.8  &  0.010    &     \\ %ATCA 
\hline
\end{tabular}
\vskip 0.1 in
\footnotesize
\raggedright
$^a$: rms noise in $\tau(\rm HCO^+)$ per $0.8\rm\,km\,s^{-1}$ channel. \\
$^b$: ATCA calibrators unless indicated as AT20G sources from Murphy et al.\,(2010) 
\end{table*}

To understand galaxy evolution, which is driven by the life cycles of stars, it is necessary to investigate 
the origin and properties of the clouds which host star formation
in a wide range of interstellar environments. 
The Magellanic System, including the Small Magellanic Cloud (SMC), Large Magellanic 
Cloud (LMC), Magellanic Bridge, Leading Arm (LA) and Magellanic Stream offer a nearby example ($\sim50$--$60\rm\,kpc$; Keller \& Wood 2006) of an
environment with interstellar conditions that sharply contrast
what we find in the Milky Way (MW). 
In particular, the low metallicity of the Magellanic System (e.g., 
$10\%$ and $45\%$ Solar metallicity respectively in the SMC and LMC; Rolleston et al.\,2002) 
imply that heating and cooling mechanisms, 
dust-to-gas ratios, and chemical abundances may 
be representative of less-evolved systems at 
high redshift.

The Magellanic System is a complex structure rich in neutral hydrogen 
(H\,{\sc i}; e.g., Putman et al.\,2003, Stanimirovi\'c et al.\,2008, Nidever et al.\,2010).
The LA, stretching out in front of the LMC and SMC, 
and the extensive, 150$^{\circ}$-long Stream 
(Nidever et al.\,2010) trailing behind them are 
understood as features of tidal  (e.g. Murai \& Fujimoto 1980, 
Yozin \& Bekki 2014) and/or ram-pressure stripping (e.g., Moore \& Davis 1994, Mastropietro et al.\,2005) 
interactions between the SMC, LMC and MW. 
Recent UV absorption measurements support the scenario that 
the Stream originated from the SMC, 
and observed metallicity enhancements
indicate some material has been
stripped from LMC (Fox et al.\,2013, 2014, Richter et al.\,2013). 
There is a wealth of evidence for star formation within Magellanic structures outside the LMC and SMC, including diffuse H$\alpha$ emission (e.g., Meaburn 1986), and massive, young stars in the LA (e.g., Casetti-Dinescu et al.\,2014) and Bridge (e.g., Demers \& Battinelli 1998), while the Stream remains starless. However, the presence and stability of cold atomic and molecular material in these extreme dynamical environments is still uncertain. 

Unfortunately, molecular hydrogen (H$_2$), the most abundant molecule in the interstellar medium (ISM),
cannot be observed directly in cold, dense environments.
Many studies trace H$_2$ with carbon monoxide (CO), which is easily 
excited at low temperatures and has strong dipole-allowed rotational transitions. 
However, both theoretical and observational studies show that CO is less effective 
at self-shielding especially at low densities (e.g., van Dishoeck \& Black 1988, Sheffer et al. 2008). 
Therefore, the important transition regions where H$_2$ begins to form 
out of the atomic medium, are poorly traced by CO (e.g., Leroy et al.\,2009).
This ``CO-dark" H$_2$ is especially prominent in low-metallicity environments 
where the abundance of dust grains, which provides the primary shielding mechanism for CO, is low (e.g., Wolfire et al.\,2010). 
Accordingly, searches for Magellanic CO emission outside of the LMC and SMC have proven difficult.
Following a detection of cold H\,{\sc i} in absorption (Kobulnicky \& Dickey 1999), 
no CO was detected in the Bridge towards source J0311-7651 (Smoker et al.\,2000).
Several CO clouds were detected in the Bridge just outside of the SMC (Muller et al.\,2003,
Mizuno et al.\,2006), but these studies targeted only regions of high far-infrared
excess, and were very expensive in observation time.

In particular, absorption by \hcop\ appears to trace H$_2$ abundance in the MW 
(e.g., Liszt et al.\,2010), 
and is therefore a promising tool for measuring molecular gas content and kinematics even in regions
where H$_2$ has recently formed (Lucas \& Liszt 1996).
Although the thermal pressure is low in such regions,
and therefore rotational transitions of molecules are
not well excited via collisions with H$_2$ (Lucas \& Liszt 1996),
absorption lines depend on column density and are
therefore strong even in low excitation.
There are no measurements of \hcop\ absorption in the Magellanic System, 
although there are two detections of H$_2$ in the Stream 
(Sembach et al.\,2001, Richter et al.\,2001), and one detection of H$_2$ in the Bridge
(Lehner et al.\,2002) from FUSE far ultra-violet (UV) absorption.
In addition, two sensitive searches for \hcop\ absorption in HVCs in the MW, 
including several Stream directions, returned only 
one tentative detection (Akeson \& Blitz 1999, Combes \& Charmandaris 2000). 

In this paper, we present the results of a search for \hcop\ absorption 
towards 9 radio continuum sources behind the Magellanic System with the Australia Telescope Compact Array (ATCA). 
Of the 9 lines of sight observed, we 
have one detection of \hcop\ absorption at the leading edge of the Magellanic Bridge, and 8 non-detections.
In Section 2 we describe the observations and data reduction, in 
Section 3 we analyze the data and present column density estimates,
and in Section 4 we discuss the results.

\section{Observations}
\label{s:obs}

We selected our targets to be radio continuum sources located behind the Magellanic System
which are bright at $3\rm\,mm$. Of the 9 targets,
6 are ATCA calibrator sources with $3\rm\,mm$ continuum flux densities $S_{3\rm\,mm}\geq0.3\rm\,Jy$, and 3 are compact
sources from the Australia Telescope 20 GHz Survey (AT20G) with $S_{3\rm\,mm}\geq0.7\rm\,Jy$ 
(Murphy et al.\,2010). 

The observations were conducted between 28 May and 01 June 2014, 
with additional observing time between 05 and 06 July 2014 granted due to poor initial observing conditions. 
We observed the \hcop (1-0) $89.188\rm\,GHz$ transition with the $3\rm\,mm$ receivers on the ATCA
in the EW $352\rm\,m$ configuration, which corresponds to a $\sim2''$ synthesized beam.
We used the CFB\,64M-32k correlator mode to cover the 
velocity\footnote{All velocities quoted in this paper 
are in the kinematic or standard LSR frame.} range from $180 -360\rm\, km\,s^{-1}$ 
within a single zoom-band. Given that all sources are compact and bright, the target sources performed as their own phase calibrators. The ATCA calibrators 0537-441 and 1921-293 were observed once every 30 minutes for bandpass calibration, Uranus was observed for 30 minutes at the end of each session for flux calibration, and pointing was corrected on each target every 20 minutes. 

All data were reduced using standard packages in MIRIAD (Sault et al.\,1995). 
Following flagging and calibration, spectral line cubes were constructed using the task INVERT
with uniform weighting and $0.2$ $\rm km\,s^{-1}$ velocity resolution. 
Final, cleaned image cubes were produced following 1000 minor iterations in CLEAN, and
all sources were unresolved. The cubes were 
Hanning-smoothed to $0.8\rm\,km\,s^{-1}$ resolution within MIRIAD. 
We then extracted the \hcop\ spectrum 
from the central pixel of each cube,\footnote{The results are consistent with extracting the spectra directly from the UV data.}
and calculated the absorption spectrum, $\exp(-\tau(\rm HCO^{+}))$, by dividing the line by the source continuum. 

Table 1 includes position and flux information for the 9 observed sources. These include:
(1) source name, (2-3) RA and Dec (J2000), (4-5) Galactic latitude and longitude ($^{\circ}$), 
(6) measured continuum flux density at $3\rm\,mm$ (Jy),  
(7) total on-source integration time (hr), 
(8) rms noise in \hcop\ optical depth, per $0.8\rm\,km\,s^{-1}$ channel, and (9) source type. 

\section{Analysis}
\label{s:analysis}

\subsection{Spectral Line Fitting}

To search for absorption signatures in the observed \hcop\ spectra, 
we used the Bayesian analysis technique developed, implemented and described 
by Allison et al. (2012). Given a likelihood function and prior distribution, this
procedure searches all parameter space using 
the \textsc{MULTINEST} Monte Carlo sampling algorithm (Feroz \& Hobson 2008) and 
calculates the resulting posterior probability distribution and evidence statistic. 
For this study, we tested for the presence of a single Gaussian function in each un-smoothed
\hcop\ absorption spectrum given the following priors:
(1) the line width is not less than the channel spacing ($0.2\rm\,km\,s^{-1}$) or 
greater than the bandwidth ($180\rm\,km\,s^{-1}$); (2) the spectral noise is equal to the
standard deviation in off-line channels; 
(3) the spectral channels are independent. 
We considered an absorption line to be ``detected" if the Bayesian evidence statistic of the Gaussian model 
is greater than the evidence statistic for a model containing no spectral lines. 

After analyzing all spectra, we detected one absorption line in the spectrum towards J0454-810. 
We display the J0454-810 spectrum with the Gaussian model in the top panel of Figure 1. 
The detection significance is described by $R=7$, where $R$ is the natural logarithm of the 
Bayesian evidence statistic for the Gaussian model divided by evidence statistic for a model containing only noise.
The line is located at \hcopvel\, has a full width at half maximum 
of \hcopwidth\ and optical depth \hcopdepth. The integrated
optical depth of the line derived from the model is, $\int \tau(\mathrm {HCO}^+)\,\mathrm{d} v = 0.44\pm0.08$, which indicates 
a $\sim5\sigma$ detection significance, and is consistent with the quoted $R$ value.

\begin{table}
\caption{Column Densities and Limits}
\centering
\begin{tabular}{lccc}
\hline
\hline 
Source & \multicolumn{1}{c}{{ $N$(H\,{\sc i})$_{\rm Mag}$    }}   &   \multicolumn{1}{c}{{$N$(\hcop)  }}  & \multicolumn{1}{c}{{$N$(H$_2$)$_{\rm Mag}$ }}    \\
                    &  ($10^{20}$\pcc)    &     ($10^{12}$\pcc)$^a$    &  ($10^{20}$\pcc)$^b$    \\
\hline	
J0056-572
& $ 0.02 \pm 0.01 $ &  $\leq 0.88 $ & $\leq 4.4 $   \\
J0102-7546
& $ 0.21 \pm 0.03 $ &  $\leq 1.34 $ & $\leq 6.7 $   \\
J0208-512
& $ 0.05 \pm 0.02 $ &  $\leq 0.76 $ & $\leq 3.8 $  \\
J0311-7651
&  $ 0.92 \pm 0.02 $ &  $\leq 0.81 $ & $\leq 4.1 $  \\
J0440-6952
&  $ 1.00 \pm 0.02 $ &  $\leq 0.68 $ & $\leq 3.4 $  \\
J0454-810
&  $ 0.42 \pm 0.03 $ &  $0.49 \pm 0.09 $ & $2.45 \pm 0.45 $   \\
J0506-6109
&  $ 0.01 \pm 0.02 $ &  $\leq 1.34 $ & $\leq 6.7 $   \\
J0530-727
&  $ 7.17 \pm 0.03 $ & $\leq 0.28 $ & $\leq 1.4 $   \\
J0637-752
&  $ 0.13 \pm 0.03 $ & $\leq 0.15 $ & $\leq 0.8 $  \\
\hline
\end{tabular}
\vskip 0.1 in
\footnotesize
\raggedright
$^a$: Assuming $4.5\rm\,km\,s^{-1}$ line width. \\
$^b$: Based on $N$(\hcop) and $X$(\hcop)$=2\times10^{-9}$.
\end{table}

As a sanity check, we note that the detected \hcop\ absorption line is present 
in the J0454-810 spectrum for: (1) each of the 4 individual days of observations,
(2) both $0.2\rm\,km\,s^{-1}$ and $0.8\rm\,km\,s^{-1}$ velocity resolutions, (3) data reduced using
different subsets of the available ATCA antennas, and (4) both linear polarizations. We are therefore confident
that the feature is not spurious. 

\begin{figure}
\begin{center}
\includegraphics[width=0.5\textwidth]{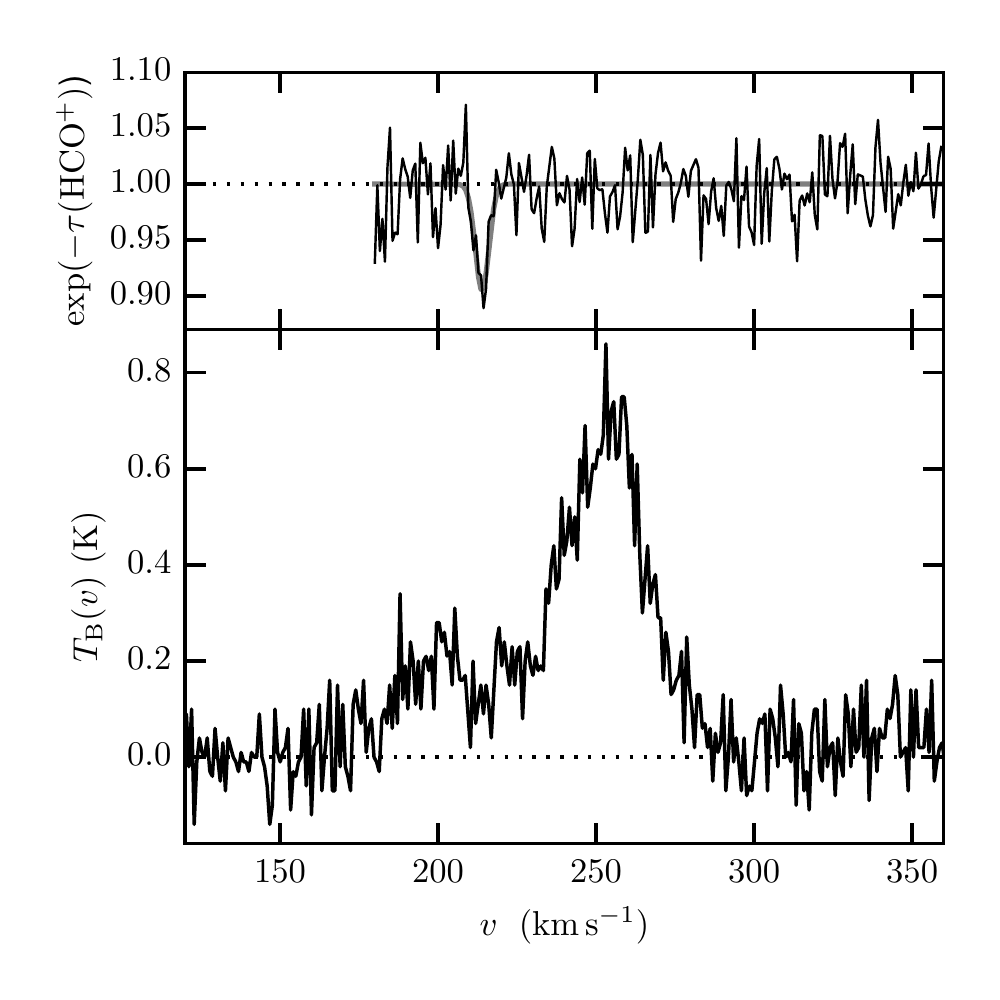}
\caption{Top: ATCA \hcop\ absorption spectrum, $\exp(-\rm \tau(HCO^{+})$), towards J0454-810. 
The Gaussian model obtained using Bayesian analysis 
(Section 3.1) is overlaid in gray. Bottom: H\,{\sc i} brightness temperature spectrum 
($T_{\rm B}(v)$) in the direction of J0454-810 from GASS (McClure-Griffiths et al.\,2009).
Two peaks appear in H\,{\sc i} emission ($v \sim200$, $260\rm\,km\,s^{-1}$),
offset from the detected \hcop\ absorption (\hcopvel).}
\end{center}
\label{fig:detection}
\end{figure}

\begin{figure*}
\begin{center}
\vspace{-20pt}
\includegraphics[width=\textwidth]{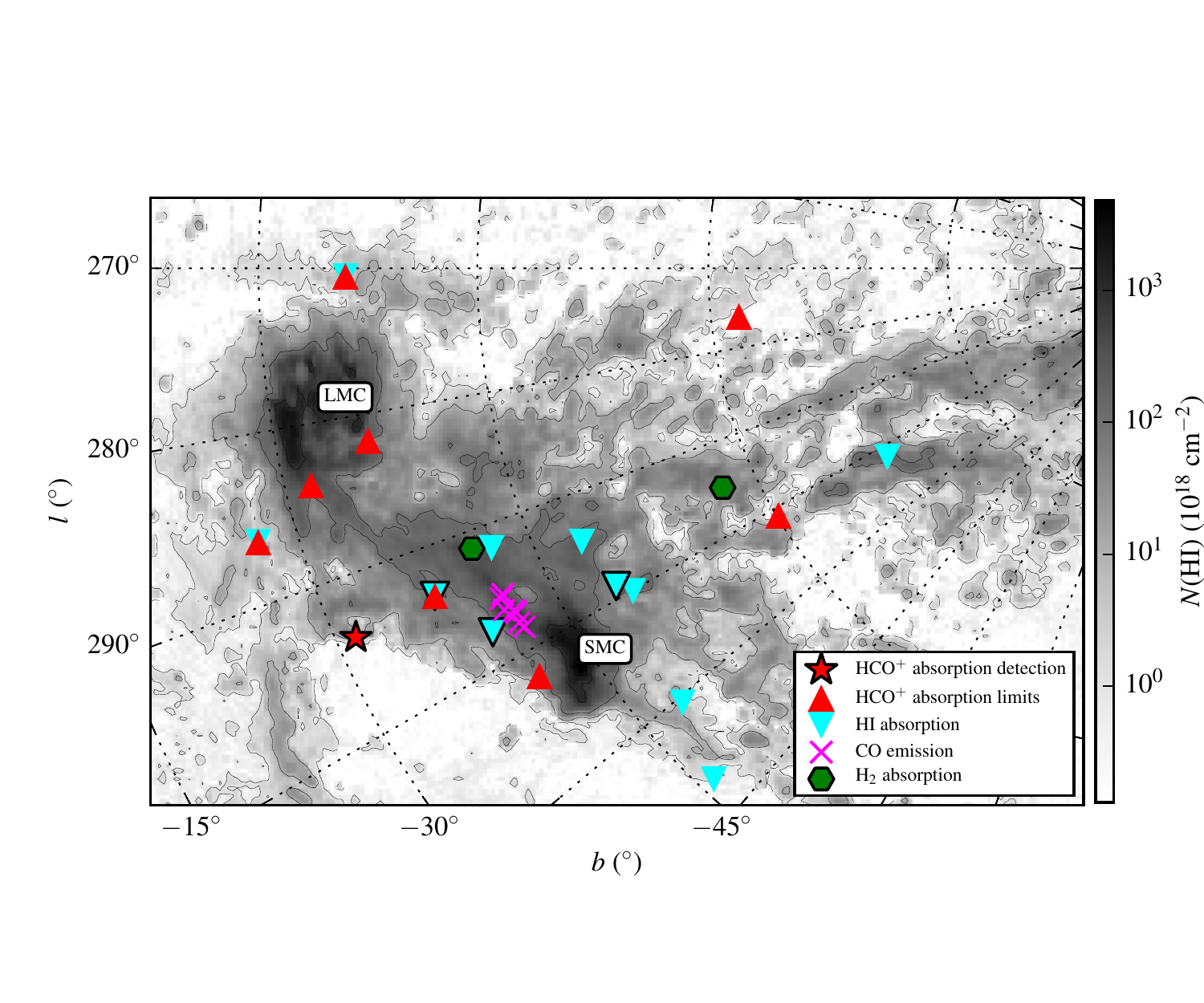}
\vspace{-60pt}
\caption{Positions of the ATCA \hcop absorption detection (red star) and limits (red triangles) overlaid on an N(H\,{\sc i}) map of the Magellanic System from Putman et al.\,(2003), where the contours correspond to $(1,10,100,1000)\times10^{18}\rm\,cm^{-2}$. We overlay H\,{\sc i} absorption (Kobulnicky \& Dickey 1999, Matthews et al.\,2009), CO emission (Muller et al.\,2003, Mizuno et al.\,2006) and H$_2$ absorption (Richter et al.\,2001, Lehner 2002) outside the LMC and SMC. Detections are distinguished from limits by solid black symbol outlines. }
\label{map}
\end{center}
\end{figure*}

\subsection{Column Densities}

Assuming excitation in equilibrium with the cosmic microwave background (Lucas \& Liszt 1996),
we calculated the \hcop\ column density, $N$(\hcop), from the J=1-0 optical depth ($\tau(\rm HCO^+)$) for 
an \hcop\ dipole moment of 3.92 Debye (Mount et al.\,2012), so that,
\begin{equation}
N(\mathrm{HCO}^+) = 1.107 \times 10^{12}\,\mathrm{cm}^{-2} \int \tau(\mathrm{HCO}^+)(v)\, \mathrm{d} v,
\end{equation}

\noindent where the line integral is expressed in $\rm km\,s^{-1}$.

For the J0454-810 detection, we integrated the Gaussian model to compute $N$(\hcop) using Equation 1. For the non-detection sight lines, we computed a limit to $N$(\hcop) using a $3\sigma$ upper limit to $\tau(\rm HCO^+)$ (with $\sigma_{\tau(\rm HCO^+)}$ in Table 1) and assumed a line width of $4.5\rm\,km\,s^{-1}$ based on the J0454-810 detection. 

From MW studies, there is a correlation between $N$(\hcop) and 
$N$(H$_2$), given by $X$(\hcop)=$N$(\hcop)/$N$(H$_2$) $\sim 2\times10^{-9}$ (e.g. Liszt et al.\,2010).
Although the correlation may be different in the lower-metallicity Magellanic System, 
we applied this factor to $N$(\hcop) to estimate the Magellanic H$_2$ column density, $N$(H$_2$)$_{\rm Mag}$. 

In addition, we extracted H\,{\sc i} emission spectra from the stray radiation-corrected 
Galactic All Sky Survey (GASS; McClure-Griffiths et al.\,2009, Kalberla et al.\,2010) for each line of sight. From these data
we computed $N$(H\,{\sc i})$_{\rm Mag}$, assuming the H\,{\sc i} is 
optically thin:
\begin{equation}
N\mathrm{(H\,{\textsc i})_{Mag}}= C_0\, \int_{180}^{360} T_{\mathrm{B}}(v)\,\mathrm{d}v,
\end{equation}

\noindent where $C_0=1.823 \times10^{18}\,\mathrm{cm^{-2}\,K^{-1}\,(km\,s^{-1})^{-1}}$ 
and $T_{\rm B}(v)$ (K) is the H\,{\sc i} brightness temperature 
integrated between radial velocities: $180\leq v \leq 360\rm\,km\,s^{-1}$. 
All column densities described above are included in Table 2.

\subsection{Estimating dust, H$_2$ and CO from \emph{Planck}}
\label{Planck}

Using available \emph{Planck} maps (version 1.20) and H\,{\sc i} emission from GASS, we can estimate limits to the Magellanic dust reddening, H$_2$ column density and CO brightness along each line of sight. 

To correct for the Galactic contribution, we computed the total Galactic H\,{\sc i} column density, $N\rm(H\,{\textsc{i}})_{MW}$, by integrating the GASS $T_{\rm B}(v)$ profile in Equation 2 for all $v \leq 180\rm\,km\,s^{-1}$.
We then converted this column density to an estimate of the reddening due to the MW, $E(B-V)_{\rm MW}$, using the results from Liszt (2014): $E(B-V)_{\rm MW} = N\rm(H\,\textsc{i})_{MW}/8.3\times10^{21}\rm\,cm^{-2}\,mag^{-1}$, which holds for $E(B-V)<0.3\rm\,mag$.
To find the total reddening along the line of sight, $E(B-V)_{\rm tot}$, we used the \emph{Planck} model of dust radiance, or total dust emission integrated over frequency, $\mathcal{R}$. 
The \emph{Planck} collaboration showed that $\mathcal{R}$ is a better 
tracer of the total column density than $353\rm\,GHz$ dust optical 
depth in diffuse, high-latitude regions, and exhibits the 
strong correlation: $E(B-V)/\mathcal{R}=(5.40\pm0.09)\times10^5$ 
(Planck Collaboration et al.\,2014). We used this correlation
to compute $E(B-V)_{\rm tot}$, and then subtracted $E(B-V)_{\rm MW}$ from $E(B-V)_{\rm tot}$ to estimate the total Magellanic contribution, $E(B-V)_{\rm Mag}$. For all lines of sight, we find $E(B-V)_{\rm Mag}\sim0.001-0.01\rm\,mag$. 

We note that the above calculations require the assumption that 
the dust to gas ratio is constant across the sampled volume. Using results from FUSE observations of the SMC and LMC, we converted $E(B-V)_{\rm Mag}$ to another estimate of $N\rm(H_2)_{Mag}$ given $N{\rm(H_2)}/E_{B-V}=6\times10^{20}\rm\,cm^{-2}\,mag^{-1}$ (Welty et al.\,2012). For all lines of sight, we find $N{\rm(H_2)_{Mag}}\sim10^{17}-10^{18}\rm\,cm^{-2}$.
Finally, we estimated the Magellanic CO brightness, $W{\rm_{CO,\,Mag}}=N{\rm(H_2)_{Mag}}/X_{\rm CO,\,Mag}$, by assuming (from observations of the SMC): $X_{\rm CO,\,Mag}=4\times10^{21}$ $\rm\,cm^{-2}\,(K\,km\,s^{-1})^{-1}$ (e.g., Bolatto et al.\,2013). For all lines of sight, we find $W{\rm_{CO,\,Mag}}\sim10^{-3}-10^{-4}\rm\,K\,km\,s^{-1}$.

\section{Discussion}
\label{s:discussion}

\subsection{Comparisons with Previous Work}

The \hcop\ absorption detection towards J0454-810 is the first of its kind 
in the Magellanic System, and suggests the presence of molecular gas
far outside of the SMC and LMC.
Of the 8 non-detections, only 2 were more sensitive and 6
were up to three times less sensitive due to compromised weather conditions.
Overall, our ATCA observations are more sensitive than previous searches
for \hcop\ absorption in diffuse, high-velocity clouds. These previous studies had typical $3\sigma$ limits to \hcop\ optical
depth of $\sim0.1-0.9$ per $0.6\rm\,km\,s^{-1}$ channels and returned no detections (Akeson \& Blitz 1999)
and one tentative and non-confirmed detection (Combes \& Charmandaris 2000) in clouds
with $N$(H\,{\sc i})$<10^{20}\rm\,cm^{-2}$. 

Our detected absorption line is wider (\hcopwidth )
than individual \hcop\ absorption lines detected in the MW ($\Delta v\sim1-2\rm\,km\,s^{-1}$; 
Lucas \& Liszt 1996), and the only previously reported high-velocity cloud detection 
($\Delta v =1.1\rm\,km\,s^{-1}$; Combes \& Charmandaris 2000). Although our measured Magellanic column density, $N\rm(HCO^{+})=(0.49\pm0.09)\times10^{12}\rm\,cm^{-2}$, is slightly
smaller than the typical column densities observed in the MW ($\sim1-10\times10^{12}\rm\,cm^{-2}$; Lucas \& Liszt 1996), it is similar to 
those observed in diffuse regions. For example, for the only sightline with $N$(H\,{\sc i})$\sim10^{20}\rm\,cm^{-2}$, 
Liszt et al.\,(2010) measure $N\rm(HCO^{+})=(0.25\pm0.08)\times10^{12}\rm\,cm^{-2}$. 

In Figure~\ref{map} we overlay the observed \hcop\ absorption coordinates 
on an $N$(H\,{\sc i}) image of the Magellanic System from the Parkes radio 
telescope (Putman et al.\,2003). 
The detection is located at the leading edge of the Magellanic Bridge.
Nearby regions contain other evidence for cold atomic and molecular gas, including H\,{\sc i} 
detected in absorption with temperatures $\sim20-70\rm\,K$ 
(Kobulnicky \& Dickey 1999, Matthews et al.\,2009), and 
CO emission close to the SMC (Muller et al.\,2003, Mizuno et al.\,2006). 
However, the detected \hcop\ absorption probes molecular gas
farther away from the LMC and SMC (about 10 kpc from the SMC's center) 
than any previous radio or mm-wave studies. 

Our estimates of $N$(H$_2$)$_{\rm Mag}$ in Table 2 ($\sim10^{20}\rm\,cm^{-2}$)
are orders of magnitude higher than both the \emph{Planck} based estimates ($\sim10^{17-18}\rm\,cm^{-2}$) and those from FUSE UV-absorption in the Stream and Bridge.
Towards Fairall 9, located at the beginning of the Stream in a region of enhanced metallicity
likely stripped from the LMC, Richter et al.\,(2013b) measured 
$\log(N\rm(H_2)_{Mag})=17.93^{+0.19}_{-0.16}\rm\,cm^{-2}$.
Towards an early-type star in the Bridge, Lehner (2002) 
measured $\log(N\rm(H_2)_{Mag})\approx15.4\rm\,cm^{-2}$. 
The discrepancies between these values and our Table 2 estimates 
suggest that the that the MW-based conversion factor $X(\rm HCO^+)$
may not be applicable in the lower-metallicity Magellanic regime.
Given that the $X(\rm CO)$ factor is expected to be up to a 
factor of 100 higher in low-metallicity regimes (e.g., Bolatto et al.\,2013), it is likely that $X(\rm HCO^+)$ 
will be similarly different in the Magellanic system. 
Additional measurements of \hcop\ and H$_2$ column densities are required to
constrain this value further.

\subsection{Origin of the Detected \hcop }

The origin and survival of the detected \hcop\ 
in the highly diffuse ($N(\rm H\,{\textsc i})=(0.42\pm0.03)\times10^{20}\rm\,cm^{-2}$) region of the Bridge is puzzling. 
The material either formed within the Bridge, or it originated from a stripping interaction with the LMC or, more likely given its proximity, the SMC. The age of the Bridge is constrained to $100-300\rm\,Myr$ by the oldest 
detected stellar populations known to have formed in-situ (e.g., Harris 2007). 
Within the Bridge, low metallicities ($\sim0.05$ Solar; Lehner 2008) indicate reduced dust grain formation rates, 
and young stellar populations (e.g., Demers \& Battinelli 1998) indicate potentially strong UV radiation fields ($10-100\times$ MW value).
These two properties, as discussed by Lehner 2002, imply that molecule formation and destruction equilibrium cannot be 
reached within the Bridge's lifetime for reasonable hydrogen volume densities (e..g, $n_H\leq100\rm\,cm^{-3}$; Lehner 2002).
Therefore, extremely high densities ($n_H\gg100\rm\,cm^{-3}$) are 
likely required in the Bridge for equilibrium molecule formation to occur.
This was also discussed by Kobulnicky \& Dickey (1999) in the 
context of cold atomic gas formation in the Bridge. 

However, non-equilibrium molecule formation may be responsible.
Simulations show that ram-pressure effects during the SMC-LMC interaction
which created the Bridge (required to explain the Bridge's observed in-situ star formation; e.g., Connors et al.\,2006, Besla 2012)
can produce high-density, star-formation-ready peaks within the diffuse medium (e.g., Mastropietro et al.\,2009).
Further simulations of ram pressure-stripping of the ISM from galaxies
have shown that stars can form diffuse, gaseous tails
when ablated gas cools and condenses in the turbulent wake of the stripping interaction (e.g., Tonnesen et al.\,2012).
In the bottom panel of Figure 1 we display the H\,{\sc i} brightness temperature 
spectrum from GASS (McClure-Griffiths et al.\,2009).
There is an offset between the two peaks in H\,{\sc i} and the location of \hcop\ absorption,
which may be indicative of a shock driven by dynamical interactions and support for molecule formation in the post-shock material.

In terms of molecule survival in this environment, dust is very important, 
as it provides shielding against the UV radiation field and acts as a 
catalyst for grain formation. Our \emph{Planck}-based 
estimates indicate very little dust and CO along the lines of sight probed by this study. 
A lack of shielding by dust or sufficiently large $N$(H\,{\sc i}) 
means that the molecular material identified by the \hcop\ detection must disperse quickly, 
and we may be observing the material in the process of being disrupted.  

Alternatively, the detection presented here may not be tracing dense molecular gas.
Comprehensive studies using the Plateau de Bure Interferometer (e.g., Lucas \& Liszt 1996, Liszt et al.\,2010)
have shown that strong absorption by \hcop, $^{12}$CO, CN, HCN, HNC and C$_2$H
can be found in surprisingly diffuse MW regions ($A_V < 1\rm\,mag$) where fully-developed molecular chemistry is
not expected to exist. Recent \emph{Herschel Space Observatory} observations have detected
similar molecular species to \hcop, including OH$^+$, in extremely diffuse gas where the fraction of H$_2$
is $<5\%$ (Indriolo et al.\,2015). 
These studies indicate that many diatomic and
polyatomic molecules can form and reach significant abundances
even when they are not well-shielded.

In support of this scenario, there is a growing supply of 
evidence for the existence of molecules and star formation in H\,{\sc i}-diffuse, dust-poor structures. 
For example, low-$N$(H\,{\sc i}) intermediate-velocity clouds
in the MW halo exhibit ubiquitous H$_2$ absorption (Richter et al.\,2003a,b).
Widespread CO emission has been detected in the ram pressure-stripped tail of 
the Norma cluster galaxy ESO 137-001, $40\rm\,kpc$ away from the disk (J\'achym et al.\,2014). 
In addition, young stellar associations in low-N(H\,{\sc i}) regions ($\sim10^{20}\rm\,cm^{-2}$)
with high H\,{\sc i} velocity dispersions ($\sim30\rm\,km\,s^{-1}$) have been detected in 
stripped tails up to $30\rm\,kpc$ away from NGC 1533 (Werk et al.\,2008). 

Overall, additional observations of molecular gas tracers at high sensitivity
and in a wide range of galactic environments are needed to understand
the formation and evolution of molecular gas at these extremes. 
\hcop\ absorption is a promising tracer of this material,
and future observations with ALMA will be able to confirm the detection
presented here, and extend the sample of available sources to probe the full 
extent of the Magellanic System and the outskirts of galaxies at higher redshifts.

\section{Summary}

Using the ATCA, we detected \hcop\ absorption in the Magellanic System 
for the first time. 
We observed 9 background continuum sources, and detected one absorption line towards J0454-810, with 
an \hcop\ column density of $N(\rm HCO^+)=(0.49\pm0.09)\times10^{12}\rm\,cm^{-2}$.
The detection is located in a diffuse (H\,{\sc i} column density of $N\rm(H\,{\textsc{i}})<10^{20}\rm\,cm^{-2}$) region at the leading edge of the Magellanic Bridge, where there is little current evidence for either dust or CO emission based on \emph{Planck} and GASS H\,{\sc i} 
observations. Despite the low $N\rm(H\,{\textsc i})$, this molecular material 
may have formed as part of a dynamical interaction during the evolution of the Bridge.  
Ultimately, higher-sensitivity searches for \hcop\ absorption 
and other molecular gas tracers across the Magellanic System are strongly needed to better 
understand the formation history and star formation potential of this 
complex, low-metallicity structure.

\acknowledgements
This work was supported by the National Science Foundation (NSF) Early Career 
Development (CAREER) Award AST-1056780. C.\,E.\,M. acknowledges 
support by the NSF Graduate Research 
Fellowship and the Wisconsin Space Grant Institution.
We thank the scheduling team at the Australia Telescope
Compact Array (ATCA) for granting additional observing time.
The ATCA is part of the Australia Telescope National Facility 
which is funded by the Commonwealth of Australia 
for operation as a National Facility managed by CSIRO.
The National Radio Astronomy Observatory is operated by Associated 
Universities, Inc. under contract with the National Science Foundation.

\bibliographystyle{apj}

\label{lastpage}
\end{document}